\documentclass[letterpaper,twocolumn,10pt]{article}
\usepackage{usenix-2020-09}
\usepackage{listings}
\usepackage[export]{adjustbox}
\usepackage{tabularx}
\usepackage{float}
\usepackage{multirow}
\usepackage{enumitem}
\usepackage{xcolor}
\usepackage{siunitx}

\microtypecontext{spacing=nonfrench}

\begin{document}

\date{}

\title{\Large \bf An Evaluation of WebAssembly and eBPF as Offloading Mechanisms \\ in the Context of Computational Storage}

\author{
{\rm Wenjun Huang}\\
German Aerospace Center \\
\href{mailto:wenjun.huang@dlr.de}{wenjun.huang@dlr.de}
\and
{\rm Marcus Paradies}\thanks{Corresponding author}\\
German Aerospace Center\\
\href{mailto:marcus.paradies@dlr.de}{marcus.paradies@dlr.de}
} 

\maketitle

\begin{abstract}
\noindent
As the volume of data that needs to be processed continues to increase, we also see renewed interests in near-data processing in the form of computational storage, with eBPF (extended Berkeley Packet Filter) being proposed as a vehicle for computation offloading.
However, discussions in this regard have so far ignored viable alternatives, and no convincing analysis has been provided.
As such, we qualitatively and quantitatively evaluated eBPF against WebAssembly, a seemingly similar technology, in the context of computation offloading.
This report presents our findings.
\end{abstract}

\section{Introduction}

Lately, a data processing paradigm called ``computational storage'' has been gaining traction.
The idea behind it is to do ``code shipping'' and process data within storage devices directly, thereby reducing data movement and
energy consumption~\cite{barbalace2021computational}.

While past research and commercial development in this direction have largely focused on specialized, ad-hoc hardware or software that tries to eke out as much performance as possible for specific types of workloads~\cite{246548,picoli2019ox,vinccon2020nkv,zhang2020scaleflux}, lately we have observed some renewed interests in enabling general computation offloading (i.e., user-programmable, dynamically loadable computation).

However, recent discussions in this direction~\cite{nvmecsspec,sniacsspec,sniaebpf} have focused almost exclusively on using eBPF~\cite{ebpf} as the offloading mechanism, ignoring other platform-independent bytecodes.
Naturally, we have some reservations about this trend, as there has yet been a convincing analysis showing that eBPF is the best option for the job.

In this work, we hope to initiate the conversation around general offloading mechanisms by qualitatively and quantitatively evaluating eBPF and WebAssembly.
Specifically, we would like to pose and answer the following questions:
\begin{itemize}[itemsep=-2pt,topsep=0pt,partopsep=0pt]
  \item What are the current states of the options?
  \item What are the advantages and disadvantages of each of the options?
  \item How can we improve these options with regards to computational storage (and near-data processing in general)?
\end{itemize}

\section{Background}

In this section, we describe briefly the various topics and building blocks relevant to our work.

\smallskip

\noindent\textbf{Near-data processing~\cite{barbalace2017s,gu2016biscuit,acharya1998active}~~}
Near-data processing (NDP) puts computations close to data to reduce data movement.
It promises a number of benefits, such as avoiding I/O bottlenecks.
It can take various forms, including, but not limited to, processing in-memory (PIM) and computational storage.

The idea of NDP is not new, and research about it can be dated back to at least 2 decades ago.
Nonetheless, we note that recently there has been a resurgence of interests in NDP, both in academia and in industry (e.g.,~\cite{nider2021case,samsungpim}).

\smallskip

\noindent\textbf{Computational Storage~\cite{barbalace2021computational,sniacsspec}~~}
Strictly speaking, computational storage is a special case of NDP that concerns only with processing data inside of storage devices. However, in practice, computational storage is not always implemented in storage (e.g., in the \emph{computational storage array (CSA)}  architecture).
In that sense, \emph{near-storage processing} is perhaps a better name for it.

Storage devices that support this kind of data processing are called \emph{computational storage devices (CSDs)}, and there are already a few generally available CSDs on the market (e.g.,~\cite{ngd,samsungcsd}).
Moreover, there are some ongoing standardization efforts as well (e.g.,~\cite{sniacsspec,nvmecsspec}).

\smallskip

\noindent\textbf{eBPF~\cite{ebpf}~~}
eBPF is an umbrella term covering numerous pieces of technology that enable running sandboxed programs inside the Linux kernel (i.e., ``kernel programming''), but we are specifically interested in the register-based eBPF bytecode here.
eBPF has a number of use cases, such as packet filtering and system monitoring with minimal overhead.
Consider the following possible workflow: users can write an I/O tracer in C and compile it into eBPF bytecode using \textit{clang}, and the bytecode can then be loaded into the kernel using \textit{libbpf}~\cite{libbpf}.

While eBPF is primarily used in the context of kernel programming, efforts have been made to enable running eBPF bytecode in a standalone
fashion~\cite{ubpf,rbpf,gershuni2019simple,genericebpf}.
However, to the best of our knowledge, there has yet been any significant usage of eBPF for general-purpose computations.

\smallskip

\noindent\textbf{WebAssembly~\cite{haas2017bringing,wasm}~~}
WebAssembly is a portable, low-level bytecode format for a stack-based VM.
It is designed for secure and efficient execution in web browsers, but recent research has shown that WebAssembly can still use some improvements in terms of security and efficiency~\cite{jangda2019not,lehmann2020everything}.

To generate WebAssembly code, tooling exists for several languages (e.g.,~\cite{rustpython} for Python and~\cite{emscripten} for C/C++).
To execute WebAssembly code outside of browsers, one can use \textit{wasmtime}, a VM with JIT and embedding support~\cite{wasmtime}, \textit{wasmer}~\cite{wasmer}, a more customizable option than \textit{wasmtime}, or \textit{GraalVM},\footnote{As
of v21.2, \textit{GraalVM}'s WebAssembly support is still ``experimental.''} a
VM with JIT and language interop support~\cite{graal}.
To interface with certain system calls (e.g., for I/O), WASI may be used~\cite{wasi}.

\section{Limitations}

Before diving into the evaluation, we would like to address some potential issues with our work.

\smallskip

\noindent\textbf{Implementation Dependence~~}
Given that both eBPF and WebAssembly are just bytecode formats at the core, any evaluation of them that goes beyond mere theoretical analysis will inevitably be implementation-dependent.
However, we believe that such evaluations still have their value.

First, having an implementation-dependent analysis is better than having no
analysis at all.
Second, from an engineering point of view, ``what is achievable today?'' is just as important as ``what is possible in the
future?''
If a certain solution is rather lacking at the moment, the question of whether we can justify the costs of improving it naturally arises.
Finally, if the analysis happens to reveal some potential improvements for the implementations, then being implementation-dependent actually helps to ``kill two birds with one stone''.

\smallskip

\noindent\textbf{Single Node~~}
We do not concern ourselves with offloading in a distributed setting, as then the focus would lean towards scheduling and coordination, which are out of the scope of this work.

\smallskip

\noindent\textbf{Safety and not Security~~}
We will discuss the safety (i.e., protection against human errors) afforded by the offloading mechanisms, but not their security (i.e., protection against attacks), as we are not security experts.

That said, we speculate that securing a WebAssembly runtime is likely easier than securing an eBPF runtime, as the former has some isolation features built in (e.g., linear memory~\cite{haas2017bringing}), but the latter relies more heavily on load-time verification~\cite{ebpf}, which not only is prone to false positives~\cite{gershuni2019simple} but also has had flaws discovered in the past (e.g.~\cite{cve-2020-8835,cve-2021-20268}).

Moreover, WebAssembly has already undergone a significant amount of scrutiny regarding its semantics and security~\cite{haas2017bringing,lehmann2020everything,watt2018mechanising,watt2019weakening,narayan2021swivel,mcfadden2018security,hilbig2021empirical}, but the same cannot yet be said of eBPF in the context of general-purpose computations.

\section{Qualitative Evaluation}

In this section, we will present our findings on the usability (i.e., the user experience) of the two offloading mechanisms.
Specifically, we will compare them along 6 axes: safety, tooling, compatibility, portability, ergonomics, and language-agnostic-ity.
A summary of the comparison is provided in Table~\ref{tab:summary} (* = poor, ** = ok, *** = good).

\begin{table}[t!]
  \centering
  \begin{tabularx}{0.45\textwidth}{|*{3}{>{\centering\arraybackslash}X|}}
    \hline
    & \textbf{eBPF} & \textbf{WebAssembly}\\
    \hline
    \textit{safety} & ? & ***\\
    \hline
    \textit{tooling} & ** & ***\\
    \hline
    \textit{compatibility} & * & **\\
    \hline
    \textit{portability} & ** & ***\\
    \hline
    \textit{ergonomics} & ** & **\\
    \hline
    \textit{language-agnostic-ity} & * & **\\
    \hline
  \end{tabularx}
  \caption{Evaluation summaries of the offloading mechanisms}
  \label{tab:summary}
  \vspace{-15pt}
\end{table}

\subsection{Safety}

As suggested above, we will not cover multi-tenant or public-facing usage here, so we only consider safety in terms of programmer mistakes -- ``what's the worst that can happen if the offloads misbehave due to bugs in them''?

For WebAssembly, the safety is pretty good, as the offload code will run in an isolated memory space~\cite{wasm_memory,haas2017bringing} and also has no access to the file system by default~\cite{wasi_principles}.
Since these properties are specified at the design level, they are not implementation-dependent.
While it is possible for WebAssembly code to hang or waste CPU cycles\footnote{For \textit{wasmtime}~\cite{wasmtime} hanging
can be avoided via the \href{https://docs.rs/wasmtime/0.29.0/wasmtime/struct.Store.html\#method.add_fuel}{``fuel'' feature}.}, it is rather unlikely for careless mistakes to wreck havoc outside of the WebAssembly runtime.

For eBPF, the situation is mostly dependent on the verifier.
The one used by the Linux kernel is very strict, banning many kinds of loops and memory accesses~\cite{kernel_ebpf_doc}.
However, using that verifier (and kernel-oriented verifiers~\cite{gershuni2019simple} in general) for data processing is not very practical, and a new verifier implementation is needed~\cite{kourtis2020safe}.
Therefore, we cannot yet say much about the safety of eBPF as an offloading mechanism.

Nonetheless, we suspect that getting eBPF runtime safety right will be a trickier task, as it relies on a specific verifier to catch undesirable behaviors.
In comparison, WebAssembly clearly specifies what is allowed at the spec level.

\subsection{Tooling}

\noindent\textbf{Development Tools~~}
For WebAssembly, a suite of tools called \textit{wabt} (The WebAssembly Binary Toolkit) is available for inspecting and transforming WebAssembly binaries~\cite{wabt}.

For eBPF, \textit{bpftool} can be used to inspect eBPF programs and maps~\cite{bpftool}.
It also has some features meant for in-kernel eBPF usage, such as loading programs and updating maps.
However, since \textit{bpftool} is inherently tied to in-kernel usage, eBPF programs cannot be inspected unless they are loaded or attached into the kernel.
For a standalone option, \textit{llvm-objdump} is a better choice~\cite{llvm_objdump}.

\smallskip

\noindent\textbf{Virtual Machines~~}
As mentioned before, WebAssembly has at least three feature-rich VM implementations: \textit{wasmtime}~\cite{wasmtime}, \textit{wasmer}~\cite{wasmer}, and \textit{GraalVM}~\cite{graal}.
While they share some common features like JIT support for common architectures, each offers a number of unique features as well.
For \textit{wasmtime}, as of v0.29:
\begin{itemize}[itemsep=-2pt,topsep=0pt,partopsep=0pt]
  \item profiling via \textit{perf} and \textit{VTune}
  \item debugging via \textit{lldb} and \textit{gdb}
  \item embedding in several host languages
  \item async host functions
  \item host functions from closures
  \item traps and backtraces
  \item custom memory allocator
  \item stack and heap limits
  \item caching of compiled binaries
  \item controlled code execution via fuel credits
  \item SIMD and threading
\end{itemize}

\medskip

\noindent
For \textit{wasmer}, as of v2.0:
\begin{itemize}[itemsep=-2pt,topsep=0pt,partopsep=0pt]
  \item embedding in several host languages
  \item ahead-of-time and cross compilation via LLVM
  \item multiple ways of storing compiler output
  \item dynamic and static functions
  \item custom heap management via \texttt{Tunables}
  \item errors and backtraces
  \item custom import resolution via \texttt{Resolver}
  \item support for many experimental features like SIMD
\end{itemize}

\medskip

\noindent
For \textit{GraalVM}, as of v21.2:
\begin{itemize}[itemsep=-2pt,topsep=0pt,partopsep=0pt]
  \item interop \& embedding for a dozen languages
  \item access to mature JVM technologies
  \item creating standalone native binaries (``native images'')
  \item debugging via Chrome DevTools and more
  \item built-in CPU \& memory profilers
  \item visualizing compilation graphs
\end{itemize}

\medskip

\noindent
eBPF, on the other hand, still has a long way to go in this department.
Both \textit{uBPF}~\cite{ubpf} and \textit{rBPF}~\cite{rbpf} have JIT support (\texttt{x86\_64} only), but they are still rather lacking when it comes to features like profiling and debugging, which is not surprising, as they are not meant for in-production usage.
In fact, the README of \textit{rBPF} clearly states that:

\begin{itemize}[itemsep=-2pt,topsep=0pt,partopsep=0pt]
  \item several features of the kernel eBPF implementation are not yet available (e.g., tail calls)
  \item it has a very simple verifier, so safety cannot be guaranteed
  \item the JIT compiler does not emit run-time memory checks, so crashes are possible
  \item a small number of eBPF instructions are not yet implemented
  \item \textit{uBPF} is mostly the same, so the above points likely hold for \textit{uBPF} as well
\end{itemize}

\subsection{Compatibility}

Compatibility here refers to the users' ability to re-use existing libraries in the ecosystem.
Any external dependency, including \textit{libc}, must be re-compiled and statically linked.
For WebAssembly, this means that the \textit{wasi-sdk}~\cite{wasisdk} is likely needed for compilation.
For example, to build \textit{zstd}~\cite{zstd}:

\begin{lstlisting}[belowskip=0pt,aboveskip=3pt]
CXX="wasi-sdk/bin/clang++" \
CC="wasi-sdk/bin/clang" \
AR="wasi-sdk/bin/ar" \
CFLAGS="--sysroot=wasi-sdk/share/wasi-sysroot" \
$(MAKE) -C zstd/lib libzstd.a
\end{lstlisting}

\noindent
For eBPF, building \textit{zstd} would fail, as \textit{libc} is not available in the absence of a WASI equivalent:

\begin{lstlisting}[belowskip=0pt,aboveskip=3pt]
CXX="clang++" CC="clang" CFLAGS="-target bpf -Wall -Werror -O2" make -C zstd/lib libzstd.a

make: Entering directory 'zstd/lib'
CC obj/conf_01069ccb7af67d6f99eda2da05bd0468/static/debug.o
CC obj/conf_01069ccb7af67d6f99eda2da05bd0468/static/entropy_common.o
In file included from common/entropy_common.c:18:
In file included from common/mem.h:24:
In file included from common/zstd_deps.h:27:
In file included from /usr/lib/llvm-12/lib/clang/12.0.0/include/limits.h:21:
/usr/include/limits.h:26:10: fatal error: 'bits/libc-header-start.h' file not found
#include <bits/libc-header-start.h>
         ^~~~~~~~~~~~~~~~~~~~~~~~~~
1 error generated.
make[1]: *** [Makefile:344: obj/conf_01069ccb7af67d6f99eda2da05bd0468/static/entropy_common.o] Error 1
make: *** [Makefile:240: libzstd.a] Error 2
make: Leaving directory 'zstd/lib'
\end{lstlisting}

\noindent
Without access to \textit{libc}, a huge swath of existing libraries would be unusable, so the compatibility of eBPF is quite poor. Additionally, \texttt{clang -target bpf}, as of v12.0.1, also does not support several kinds of numeric operations.
For example, \texttt{i64} divisions:

\begin{lstlisting}[belowskip=0pt,aboveskip=3pt]
Error: Unsupport signed division for DAG: 0x556059b1c260: i64 = sdiv 0x556059b1ef90, 0x556059b1bf20Please convert to unsigned div/mod.
fatal error: error in backend: Cannot select: 0x556059b1c260: i64 = sdiv 0x556059b1ef90, 0x556059b1bf20
  0x556059b1ef90: i64 = add 0x556059b1ef28, Constant:i64<2>
    0x556059b1ef28: i64 = add 0x556059b1bbe0, 0x556059b1bd18
      0x556059b1bbe0: i64 = mul 0x556059b1bc48, Constant:i64<7>
        0x556059b1bc48: i64 = mul 0x556059b1beb8, 0x556059b1bff0
          0x556059b1beb8: i64,ch = CopyFromReg 0x556059a67458, Register:i64 %3
            0x556059b1bf88: i64 = Register %3
          0x556059b1bff0: i64,ch = CopyFromReg 0x556059a67458, Register:i64 %7
            0x556059b1c0c0: i64 = Register %7
        0x556059b1b7d0: i64 = Constant<7>
      0x556059b1bd18: i64 = mul nsw 0x556059b1bb10, 0x556059b1ba40
        0x556059b1bb10: i64,ch = CopyFromReg 0x556059a67458, Register:i64 %5
          0x556059b1b700: i64 = Register %5
        0x556059b1ba40: i64,ch = CopyFromReg 0x556059a67458, Register:i64 %8
          0x556059b1bcb0: i64 = Register %8
    0x556059b1b8a0: i64 = Constant<2>
  0x556059b1bf20: i64 = mul nsw 0x556059b1bb10, 0x556059b1be50
    0x556059b1bb10: i64,ch = CopyFromReg 0x556059a67458, Register:i64 %5
      0x556059b1b700: i64 = Register %5
    0x556059b1be50: i64,ch = CopyFromReg 0x556059a67458, Register:i64 %2
      0x556059b1bde8: i64 = Register %2
\end{lstlisting}

\noindent
and floating point operations (passing \texttt{-fno-builtin} did not help here):

\begin{lstlisting}[belowskip=0pt,aboveskip=3pt]
4-pi.c:1:8: error: A call to built-in function '__muldf3' is not supported.
double pi_digit(unsigned long n_th) {
       ^
4-pi.c:1:8: error: A call to built-in function '__adddf3' is not supported.
4-pi.c:1:8: error: A call to built-in function '__subdf3' is not supported.
4-pi.c:1:8: error: A call to built-in function '__muldf3' is not supported.
4-pi.c:1:8: error: A call to built-in function '__ltdf2' is not supported.
4-pi.c:1:8: error: A call to built-in function '__adddf3' is not supported.
4-pi.c:1:8: error: A call to built-in function '__adddf3' is not supported.
4-pi.c:1:8: error: A call to built-in function '__muldf3' is not supported.
4-pi.c:1:8: error: A call to built-in function '__muldf3' is not supported.
4-pi.c:1:8: error: A call to built-in function '__adddf3' is not supported.
4-pi.c:1:8: error: A call to built-in function '__muldf3' is not supported.
4-pi.c:1:8: error: A call to built-in function '__muldf3' is not supported.
4-pi.c:1:8: error: A call to built-in function '__adddf3' is not supported.
4-pi.c:1:8: error: A call to built-in function '__adddf3' is not supported.
\end{lstlisting}

\noindent
The lack of support for these operations is not due to \textit{clang}.
Rather, the issue stems from the fact that eBPF lacks the appropriate instructions for \textit{clang} to emit~\cite{kernel_ebpf_doc,ebpf_unofficial_spec}.
As such, we do not see how these operations can be used by eBPF offload programs unless eBPF is given significant extensions.

Of course, the compatibility of WebAssembly is not perfect either -- since WASI does not yet provide functions like \texttt{fork()} and
\texttt{pthread\_*()}~\cite{wasi_api}, certain existing libraries would not build (e.g., the multi-threaded version of \textit{zstd}), but the situation is arguably better than eBPF.\footnote{Furthermore, as WebAssembly gains more traction, we expect to see more libraries providing official WebAssembly support. \href{https://blog.tensorflow.org/2020/03/introducing-webassembly-backend-for-tensorflow-js.html}{TensorFlow} would be one example.
For now, the same cannot be said of eBPF.}

\subsection{Portability}

Since eBPF and WebAssembly are bytecodes designed to be executed by virtual machines, they both have decent portability in theory (i.e., users can compile their code once and reuse the built binaries everywhere).

However, in practice, eBPF is at a slight disadvantage on this front because it is not endianness-independent.
The LLVM static compiler \texttt{llc} 12.0.1 gives:

\begin{lstlisting}[belowskip=0pt,aboveskip=3pt]
Registered Targets:
  ...
  bpf        - BPF (host endian)
  bpfeb      - BPF (big endian)
  bpfel      - BPF (little endian)
  ...
  wasm32     - WebAssembly 32-bit
  wasm64     - WebAssembly 64-bit
  ...
\end{lstlisting}

\noindent
For many use cases, the host endianness is not that relevant, as they do not involve parsing and interpreting data at the byte level.
However, this ``parsing'' task is arguably quite common during data processing (e.g., reading numbers out of a file to perform calculations), so eBPF's endianness-dependence will likely affect the portability of many offload binaries.

For WebAssembly, little-endianness is always used~\cite{wasm}, and it is up to the runtime to handle the conversion if the host architecture is big-endian~\cite{wasm_portability}, so there is no endianness-related portability concern.\footnote{This should not be a performance concern either, as modern architectures often provide instructions for handling endianness (e.g., \texttt{MOVBE} from x86).} 
Moreover, the difference between \texttt{wasm32} and \texttt{wasm64} is mostly about the largest memory size supported~\cite{wasm64}, which can be transparently handled by the runtime and hence should not lead to any difference in program behavior.

\subsection{Ergonomics}

For this section, we will take a look at the development processes of the two offloading mechanisms and see if they are easy and straightforward.

\subsubsection{Compilation}

\noindent\textbf{WebAssembly~~}
As mentioned before, it is possible to generate WebAssembly code from several source languages.
To have a direct comparison with eBPF, we will focus on C here.
From C code, there are at least 2 ways to compile to WebAssembly with \textit{libc} support: \textit{Emscripten}~\cite{emscripten} and
\textit{wasi-sdk}~\cite{wasisdk}.
With \textit{Emscripten}:

\begin{lstlisting}[belowskip=0pt,aboveskip=3pt]
emcc \
-Wall -O3 -flto \
--no-entry -Wl,--export-all \
-o code.wasm code.c
\end{lstlisting}

\noindent
With \textit{wasi-sdk}:

\begin{lstlisting}[belowskip=0pt,aboveskip=3pt]
wasi-sdk/bin/clang \
--sysroot=wasi-sdk/share/wasi-sysroot \
-Wall -O3 -flto -nostartfiles \
-Wl,--no-entry -Wl,--export-all \
-o code.wasm code.c
\end{lstlisting}

\noindent
\texttt{-flto} enables link-time optimization (LTO),
\texttt{-nostartfiles} disables the system \textit{libc},
\texttt{-Wl,--no-entry} allows the absence of \texttt{main()}, and
\texttt{-Wl,--export-all} makes sure that all functions, including unused
ones, are emitted.

After compilation, \textit{wasm-opt} from the \textit{binaryen} toolchain~\cite{binaryen} may be used to further reduce code size:
{\small\texttt{wasm-opt -Oz code.wasm -o code.wasm}}.

\medskip

\noindent\textbf{eBPF~~}
For eBPF, both of the two mainstream compilers, \textit{gcc} and \textit{clang}, may be used.
We will stick with \textit{clang} here since it is the more popular and mature option: {\small\texttt{clang -target bpf -Wall -O2 -o code.o -c code.c}}

Compiling to eBPF appears to be much simpler than compiling to WebAssembly.
However, if we are actually compiling for in-kernel usage, then some additional \texttt{-I} flags are likely needed to access kernel types.

\subsubsection{Bytecode Inspection}

\noindent\textbf{WebAssembly~~}
A tool called \textit{wasm2wat} is included in \textit{wabt}~\cite{wabt} for the purpose of converting WebAssembly binaries to an official, human-readable text format\footnote{\textit{wasmtime} supports executing this text format directly.}~\cite{wasm_text}.
Here is an example of this text format:

\begin{lstlisting}[belowskip=0pt,aboveskip=3pt]
(module
  (func (export "addTwo") (param i32 i32) (result i32)
    local.get 0
    local.get 1
    i32.add))
\end{lstlisting}

\noindent
One can also disassemble WebAssembly binaries using \textit{wasm-objdump}:

\begin{lstlisting}[belowskip=0pt,aboveskip=3pt]
wasm-objdump add_one.wasm

add_one.wasm:   file format wasm 0x1

Code Disassembly:

0001bc func[0] <__wasm_call_ctors>:
 0001bd: 01                         | nop
 0001be: 0b                         | end
0001c0 func[1] <malloc>:
 0001c1: 20 00                      | local.get 0
 0001c3: 10 02                      | call 2
 0001c5: 0b                         | end
0001c8 func[2]:
 0001c9: 0b 7f                      | local[0..10] type=i32
 0001cb: 23 00                      | global.get 0
 0001cd: 41 10                      | i32.const 16
 0001cf: 6b                         | i32.sub
 0001d0: 22 0a                      | local.tee 10
 0001d2: 24 00                      | global.set 0
 0001d4: 02 40                      | block
 0001d6: 41 a0 08                   |   i32.const 1056
\end{lstlisting}

\smallskip

\noindent\textbf{eBPF~~}
Although eBPF does not yet have an official textual format,\footnote{Both \textit{rBPF} and \textit{uBPF} provide (dis)assemblers for a textual assembly format, but it is not an official format.} if the eBPF binary of interest was compiled with debugging information, it is possible to view the disassembled binary with interleaved source code:

\begin{lstlisting}[belowskip=-10pt,aboveskip=3pt]
llvm-objdump -S --no-show-raw-insn bpf.o

bpf.o:  file format elf64-bpf

Disassembly of section fentry/blk_account_io_start:

0000000000000000 <blk_account_io_start>:
; int BPF_PROG(blk_account_io_start, struct request *req)
       0:       r8 = *(u64 *)(r1 + 0)
       1:       *(u64 *)(r10 - 8) = r8
       2:       r1 = 0
;   struct req_info info = { 0 };
       3:       *(u64 *)(r10 - 16) = r1
       4:       *(u64 *)(r10 - 24) = r1
       5:       *(u64 *)(r10 - 32) = r1
       6:       *(u64 *)(r10 - 40) = r1
       7:       *(u64 *)(r10 - 48) = r1
       8:       *(u64 *)(r10 - 56) = r1
       9:       *(u64 *)(r10 - 64) = r1
      10:       *(u64 *)(r10 - 72) = r1
      11:       *(u64 *)(r10 - 80) = r1
      12:       *(u64 *)(r10 - 88) = r1
      13:       *(u64 *)(r10 - 96) = r1
      14:       *(u64 *)(r10 - 104) = r1
      15:       *(u64 *)(r10 - 112) = r1
      16:       *(u64 *)(r10 - 120) = r1
      17:       *(u64 *)(r10 - 128) = r1
      18:       *(u64 *)(r10 - 136) = r1
;   u64 t = bpf_ktime_get_ns();
      19:       call 5
      20:       r6 = r0

\end{lstlisting}

\subsection{Language-agnostic-ity}

For WebAssembly, a variety of source languages can be used in theory~\cite{wasm}.
However, in reality, the source language has to support exporting \texttt{malloc()}, which is needed to allocate WebAssembly buffers and pass data between the host and the WebAssembly VM.
That said, this situation will change when the ``module linking'' feature is stabilized~\cite{wasm_module_linking}, as then it will be possible for the host to link needed functions, and more source languages may be used.

For eBPF, the only supported source language at the moment is C.
While it is possible to use other source languages in theory~\cite{rust_to_ebpf}, significant efforts are required from the user.

\section{Quantitative Evaluation}

In this section, we will describe how we evaluated the performance of the two offloading mechanisms.
We will also present the evaluation results.

\subsection{Setup}

\noindent
The evaluation was carried out on two platforms, an AMD workstation (Ryzen Threadripper Pro 3975WX~\cite{amd3975} (\texttt{x64}), 128 GiB of RDIMM DDR4 ECC DRAM (3200 MHz), Linux kernel 5.13.13)  and an AWS c6g.xlarge instance (4 vCPU from AWS Graviton2 (\texttt{arm64}), 8 GiB DRAM, Linux kernel 5.11.0) and for the configurations as summarized in Table~\ref{eval:configs}.



\begin{table}[h!]
  \centering
  \begin{tabularx}{0.45\textwidth}{|c|>{\centering\arraybackslash}X|}
    \hline
    \multirow{5}{*}{\textbf{VM}} & \textit{wasmtime} v0.29.0~\cite{wasmtime}\\
    \cline{2-2}
    \  & \textit{wasmer} v2.0.0~\cite{wasmer}\\
    \cline{2-2}
    \  & \textit{GraalVM} v21.2.0~\cite{graal}\\
    \cline{2-2}
    \  & \textit{rBPF} @ 6c524b3~\cite{rbpf}\\
    \cline{2-2}
    \  & \textit{uBPF} @ 9eb26b4~\cite{ubpf}\\
    \hline
    \multirow{3}{*}{\textbf{toolchain}} & \textit{llvm} / \textit{clang} 12.0.1 (WX)
     \& 12.0.0 (AWS)\\
     \cline{2-2}
    \  & \textit{wabt} 1.0.24 (WX) \& 1.0.20 (AWS)~\cite{wabt}\\
    \cline{2-2}
    \  & \textit{rustc} 1.57.0-nightly (e30b68353 2021-09-05)\\
    \hline
  \end{tabularx}
  \caption{Evaluation configurations}
  \label{eval:configs}
\end{table}

\noindent
Since there is currently no \textit{libc} support for eBPF, our evaluation consisted only of simple arithmetic programs that are written in C:
\begin{itemize}[itemsep=-2pt,topsep=0pt,partopsep=0pt]
  \item a baseline dummy program that simply returns 1
  \item fibonacci calculation
  \item integer summing
  \item Erd{\H o}s prime counting
  \item multi-factorial calculation
\end{itemize}

\smallskip

\noindent
For our evaluation, \texttt{clang -Oz} was used to compile to all
targets,\footnote{Of course, there were additional flags involved. Please check
the \texttt{Makefile}s in the repo.} and \textit{wasm-opt} was used to
further optimize WebAssembly binaries after \textit{clang} finishes.
Moreover, the following VM configurations were evaluated:

\smallskip

\begin{itemize}[itemsep=-2pt,topsep=0pt,partopsep=0pt]
  \item {\textit{graal}}: for running WASM binaries on \href{https://www.graalvm.org}{GraalVM} via a native image
  \item {\textit{rbpf}}: for running patched (see next section) eBPF binaries on \href{https://github.com/qmonnet/rbpf}{rBPF}
  \item {\textit{ubpf}}: for running patched (see next section) eBPF binaries on \href{https://github.com/iovisor/ubpf}{uBPF}
  \item {\textit{wasmer}}: for running WASM binaries on \href{https://github.com/wasmerio/wasmer}{wasmer}
  \item {\textit{wasmtime}}: for running WASM binaries on \href{https://github.com/bytecodealliance/wasmtime}{wasmtime}
  \item {\textit{wasmtime\_cached}}: like {\textit{wasmtime}} but with caching enabled
\end{itemize}

\smallskip

\noindent
In addition to the WebAssembly VMs and the eBPF VMs, the native binaries were also benchmarked\footnote{The native binaries were loaded using \texttt{dlopen()} and the functions were invoked via \texttt{dlsym()}, so that the native binaries were used in a way that is similar to how eBPF and WebAssembly binaries were used.} so that they may serve as a baseline.
The performance metrics shown below were collected using our own scripts and GNU \textit{time} v1.9~\cite{gnutime} over 10 consecutive runs.

Note that, since \textit{rBPF} and \textit{uBPF} do not currently provide JIT implementations for \texttt{arm64}, we did not evaluate them on that architecture.
The same applies for \textit{wasmer}, which was not able to compile the WebAssembly code on \texttt{arm64} due to LLVM errors.
Thus, the tables below will show ``NA'' for the relevant cells.

\subsection{Patching eBPF Binaries}

eBPF programs cannot allocate memory (either statically or dynamically) or call non-helper functions (i.e., functions that are not registered with the VM).
To make things work,\footnote{In our case, that means to have \textit{some} support for function calls and static memory.} the binaries produced directly by \textit{clang} had to be disassembled and manually patched for our evaluation:\footnote{See our repository for more information and examples.}

\smallskip

\begin{enumerate}[itemsep=-2pt,topsep=0pt,partopsep=0pt]
  \item in the original code file, add a pointer parameter to all functions that need external memory, which shall be accessed through the said pointer
  \item compile the code file and disassemble the binary file produced
  \item create a plaintext file
  \item put the expected output of the program on the first line; this should be a {\texttt{u64}} integer
  \item put the size of the needed static memory on the second line; this should be in unit of bytes and can be {\texttt{0}}
  \item paste the disassembled code after the second line, with the entry point pasted first and all other functions pasted afterwards
  \item replace all {\texttt{call}} instructions and all {\texttt{exit}} instructions of the callees with {\texttt{ja}} instructions to simulate function calls
  \item the address of the external memory will be placed in register {\texttt{r1}}, so make sure you do not mess with {\texttt{r1}} at the beginning
  \item make sure the file ends with an {\texttt{exit}} instruction
\end{enumerate}

\smallskip

\noindent
We will also surmise that it is unlikely for this patching process to be automated post-compilation, as \textit{clang} currently emits all eBPF function calls as \texttt{call -1}, making it impossible to stitch assembly blocks together without source information.


\begin{table*}[t]
  \begin{tabularx}{1.0\textwidth}{|c|*{5}{>{\centering\arraybackslash}X|}}
    \hline
    \textbf{program} & \textbf{VM} & \textbf{rel. avg. max RSS} & \textbf{rel. avg. latencies}
    & \textbf{rel. avg. times} & \textbf{rel. avg. times}\\
    \hline
    \multirow{6}{*}{\textit{dummy}} & \textit{wasmtime} & 3.85 & 92.0 & 4.37 & 2.71\\
    \cline{2-6}
    \  & \textit{wasmtime\_cached} & 3.52 & 80.6 & 4.25 & 2.59\\
    \cline{2-6}
    \  & \textit{graal} & 10.4 & 25.5 & 20.4 & 2.87\\
    \cline{2-6}
    \  & \textit{wasmer} & 16.0 & 267.0 & 4.63 & 7.31\\
    \cline{2-6}
    \  & \textit{rbpf} & 1.11 & 1.59 & 0.0972 & 0.852\\
    \cline{2-6}
    \  & \textit{ubpf} & 1.10 & 3.23 & 0.0887 & 1.11\\
    \hline
    \multirow{6}{*}{\textit{fibonacci}} & \textit{wasmtime} & 3.85 & 94.0 & 3.77 & 2.89\\
    \cline{2-6}
    \  & \textit{wasmtime\_cached} & 3.52 & 79.2 & 4.10 & 2.62\\
    \cline{2-6}
    \  & \textit{graal} & 10.4 & 27.9 & 34.1 & 3.22\\
    \cline{2-6}
    \  & \textit{wasmer} & 16.8 & 283.0 & 3.82 & 8.00\\
    \cline{2-6}
    \  & \textit{rbpf} & 1.09 & 2.53 & 0.185 & 0.962\\
    \cline{2-6}
    \  & \textit{ubpf} & 1.11 & 3.83 & 0.160 & 1.07\\
    \hline
    \multirow{6}{*}{\textit{integer summing}} & \textit{wasmtime} & 3.82 & 79.7 & 1.99 & 2.17\\
    \cline{2-6}
    \  & \textit{wasmtime\_cached} & 3.48 & 64.5 & 1.99 & 2.10\\
    \cline{2-6}
    \  & \textit{graal} & 21.4 & 18.7 & 12.5 & 10.6\\
    \cline{2-6}
    \  & \textit{wasmer} & 16.6 & 223.0 & 1.06 & 2.35\\
    \cline{2-6}
    \  & \textit{rbpf} & 1.09 & 2.26 & 2.22 & 1.97\\
    \cline{2-6}
    \  & \textit{ubpf} & 1.09 & 2.97 & 1.98 & 1.79\\
    \hline
    \multirow{6}{*}{\textit{prime counting}} & \textit{wasmtime} & 2.81 & 77.7 & 0.969 & 1.55\\
    \cline{2-6}
    \  & \textit{wasmtime\_cached} & 2.48 & 58.8 & 0.978 & 1.43\\
    \cline{2-6}
    \  & \textit{graal} & 39.1 & 26.1 & 36.8 & 27.2\\
    \cline{2-6}
    \  & \textit{wasmer} & 12.3 & 297.0 & 0.814 & 3.34\\
    \cline{2-6}
    \  & \textit{rbpf} & 1.02 & 4.73 & 0.907 & 0.935\\
    \cline{2-6}
    \  & \textit{ubpf} & 1.05 & 6.71 & 0.942 & 0.999\\
    \hline
    \multirow{6}{*}{\textit{multi-factorial}} & \textit{wasmtime} & 3.80 & 78.7 & 3.55 & 2.58\\
    \cline{2-6}
    \  & \textit{wasmtime\_cached} & 3.47 & 68.6 & 3.32 & 2.45\\
    \cline{2-6}
    \  & \textit{graal} & 10.2 & 22.8 & 27.1 & 2.91\\
    \cline{2-6}
    \  & \textit{wasmer} & 16.2 & 238.0 & 3.93 & 7.30\\
    \cline{2-6}
    \  & \textit{rbpf} & 1.08 & 2.12 & 0.171 & 0.997\\
    \cline{2-6}
    \  & \textit{ubpf} & 1.07 & 3.14 & 0.140 & 1.03\\
    \hline
  \end{tabularx}
  \vspace{-10pt}
  \caption{Evaluation results for all \texttt{(program,VM)} combinations from the AMD workstation.
  \textbf{Col. 3}: max resident set sizes (bytes).
  \textbf{Col. 4}: startup latencies (ms).
  \textbf{Col. 5}: execution times (ms).
  \textbf{Col. 6}: total running times (ms).
  All values are 10-run averages and are relative to those of native binaries (= 1).}
  \label{x64_table}
  \vspace{-10pt}
\end{table*}

\begin{table*}[t]
  \begin{tabularx}{1.0\textwidth}{|c|*{5}{>{\centering\arraybackslash}X|}}
    \hline
    \textbf{program} & \textbf{VM} & \textbf{rel. avg. max RSS} & \textbf{rel. avg. latencies}
    & \textbf{rel. avg. times} & \textbf{rel. avg. times}\\
    \hline
    \multirow{6}{*}{\textit{dummy}} & \textit{wasmtime} & 3.58 & 48.3 & 51.6 & 1.84\\
    \cline{2-6}
    \  & \textit{wasmtime\_cached} & 3.28 & 29.2 & 63.1 & 1.65\\
    \cline{2-6}
    \  & \textit{graal} & 13.8 & 40.4 & 233.0 & 3.60\\
    \cline{2-6}
    \  & \textit{wasmer} & NA & NA & NA & NA\\
    \cline{2-6}
    \  & \textit{rbpf} & NA & NA & NA & NA\\
    \cline{2-6}
    \  & \textit{ubpf} & NA & NA & NA & NA\\
    \hline
    \multirow{6}{*}{\textit{fibonacci}} & \textit{wasmtime} & 3.61 & 39.1 & 27.8 & 1.80\\
    \cline{2-6}
    \  & \textit{wasmtime\_cached} & 3.28 & 24.9 & 31.1 & 1.63\\
    \cline{2-6}
    \  & \textit{graal} & 13.8 & 34.7 & 241.0 & 3.58\\
    \cline{2-6}
    \  & \textit{wasmer} & NA & NA & NA & NA\\
    \cline{2-6}
    \  & \textit{rbpf} & NA & NA & NA & NA\\
    \cline{2-6}
    \  & \textit{ubpf} & NA & NA & NA & NA\\
    \hline
    \multirow{6}{*}{\textit{integer summing}} & \textit{wasmtime} & 3.53 & 39.8 & 1.24 & 1.30\\
    \cline{2-6}
    \  & \textit{wasmtime\_cached} & 7.84 & 26.4 & 1.24 & 1.31\\
    \cline{2-6}
    \  & \textit{graal} & 28.2 & 37.1 & 11.5 & 10.7\\
    \cline{2-6}
    \  & \textit{wasmer} & NA & NA & NA & NA\\
    \cline{2-6}
    \  & \textit{rbpf} & NA & NA & NA & NA\\
    \cline{2-6}
    \  & \textit{ubpf} & NA & NA & NA & NA\\
    \hline
    \multirow{6}{*}{\textit{prime counting}} & \textit{wasmtime} & 2.33 & 49.2 & 1.27 & 1.47\\
    \cline{2-6}
    \  & \textit{wasmtime\_cached} & 3.44 & 23.6 & 1.27 & 1.40\\
    \cline{2-6}
    \  & \textit{graal} & 48.0 & 45.3 & 48.1 & 38.1\\
    \cline{2-6}
    \  & \textit{wasmer} & NA & NA & NA & NA\\
    \cline{2-6}
    \  & \textit{rbpf} & NA & NA & NA & NA\\
    \cline{2-6}
    \  & \textit{ubpf} & NA & NA & NA & NA\\
    \hline
    \multirow{6}{*}{\textit{multi-factorial}} & \textit{wasmtime} & 3.55 & 38.1 & 32.3 & 1.80\\
    \cline{2-6}
    \  & \textit{wasmtime\_cached} & 3.24 & 23.7 & 37.2 & 1.67\\
    \cline{2-6}
    \  & \textit{graal} & 13.7 & 34.1 & 243.0 & 3.57\\
    \cline{2-6}
    \  & \textit{wasmer} & NA & NA & NA & NA\\
    \cline{2-6}
    \  & \textit{rbpf} & NA & NA & NA & NA\\
    \cline{2-6}
    \  & \textit{ubpf} & NA & NA & NA & NA\\
    \hline
  \end{tabularx}
  \vspace{-10pt}
  \caption{Evaluation results for all \texttt{(program,VM)} combinations from AWS c6g.xlarge.
  \textbf{Col. 3}: max resident set sizes (bytes).
  \textbf{Col. 4}: startup latencies (ms).
  \textbf{Col. 5}: execution times (ms).
  \textbf{Col. 6}: total running times (ms).
  All values are 10-run averages and are relative to those of native binaries (= 1).
  \textbf{NA} is shown for failed configurations.}
  \label{arm64_table}
  \vspace{-10pt}
\end{table*}


\subsection{Runtime Memory Footprint}

As the environments for computational storage and near-data processing tend to be resource-constrained, the memory footprints of the VMs are naturally of interest here.\footnote{It is worth noting that neither WebAssembly nor eBPF
has GC at the moment. For the former, a GC proposal is in early-stage
development~\cite{wasm_gc}, and the current approach to reclaim resources is
to use short-lived, one-off execution contexts, which admittedly are
inadequate for long-running programs.
For the latter, since in-kernel usage does not support \textit{libc} and therefore memory allocation, how memory management will be handled in a data processing context is yet to be seen.}
Thus, we measured the memory usage of the VMs for each of the programs mentioned before.

\begin{figure}[t!]
  \includegraphics[width=.99\columnwidth]{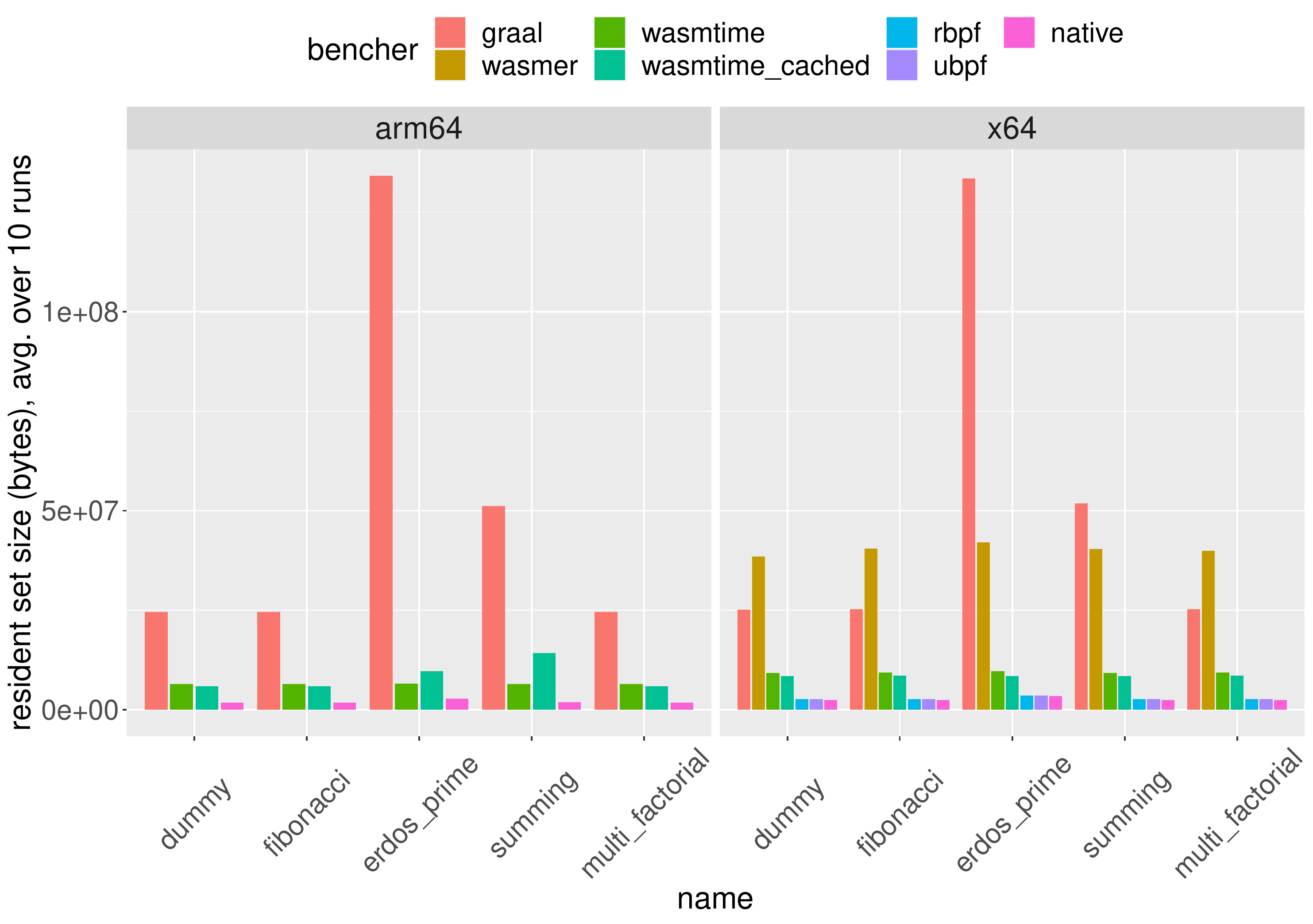}
  \vspace{-10pt}
  \caption{10-run averages of max resident set sizes (bytes) of all \texttt{(program,VM)} combinations, absolute values}
  \label{rss_bar}
  \vspace{-15pt}
\end{figure}

\begin{figure}[t!]
  \includegraphics[width=0.99\columnwidth]{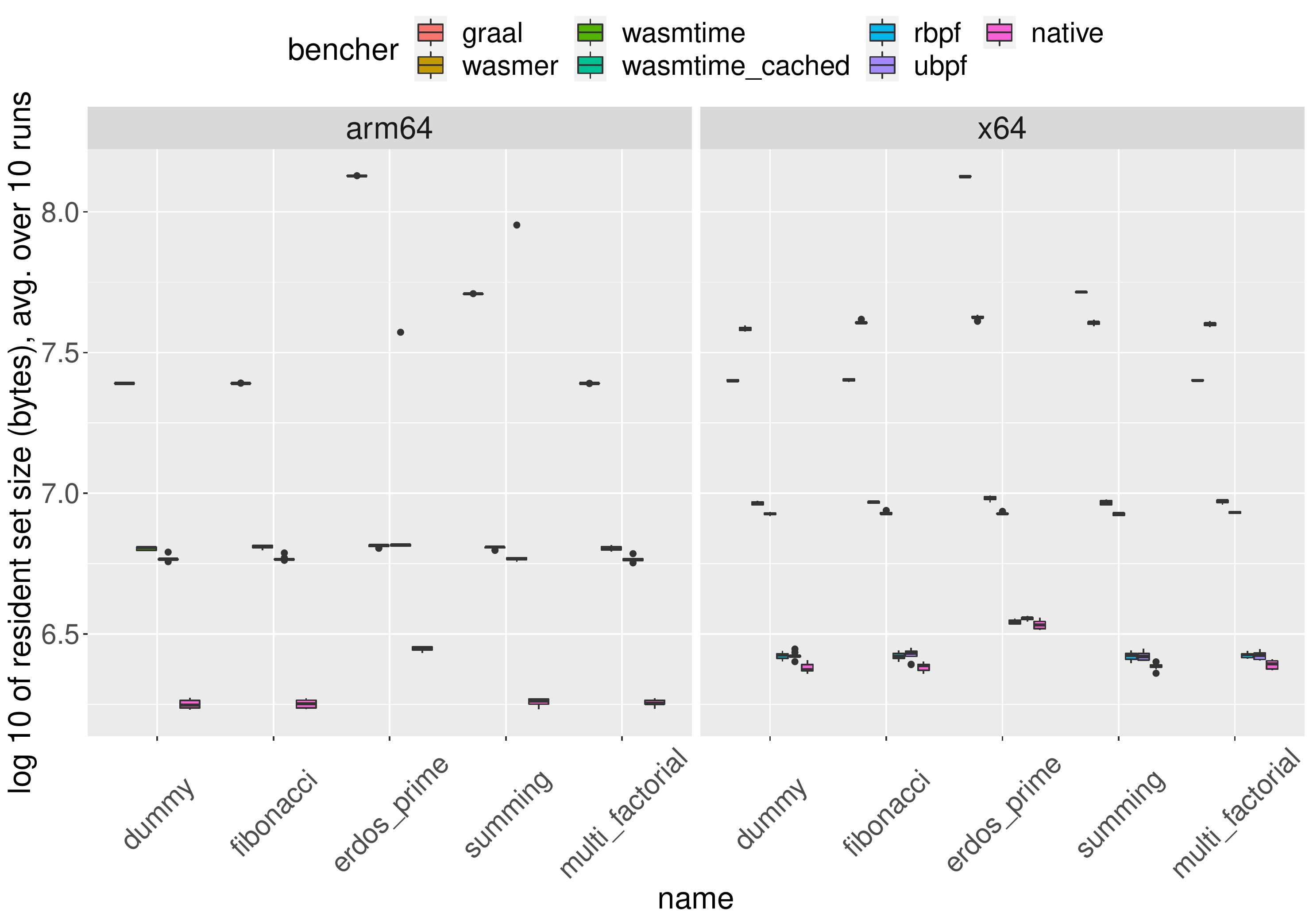}
  \vspace{-10pt}
  \caption{max resident set sizes (bytes) of all \texttt{(program,VM)} combinations, absolute values in \texttt{log10} scale}
  \label{rss_box}
\end{figure}

From Tables \ref{x64_table} and \ref{arm64_table} (column 3) and Figures \ref{rss_bar} and \ref{rss_box}, we can see that on \texttt{x64} \textit{rBPF} and \textit{uBPF} were able to achieve near-native performance when it comes to max resident set sizes (RSS).
Next comes \textit{wasmtime}, whose max RSS values were 3$\sim$4 times those of native binaries on both architectures.

The much higher max RSS values of \textit{wasmer} on \texttt{x64} were likely due to its usage of LLVM for ahead-of-time compilation (AOT).
As for \textit{GraalVM} native images, we do not have sufficient knowledge of their inner workings to comment, but we speculate that its support for language interop would require a heavier runtime, which might have contributed to the high max RSS values.

From Figure~\ref{rss_box}, we can see that the max RSS values were fairly consistent across runs, even though a few outliers do exist.
However, it is worth noting that max RSS values are \textit{max} values -- the VMs most likely do not need this much memory for their entire lifetimes.

\subsection{Runtime Startup Latency}

Generally, runtime startup latencies do not matter, as they are one-off costs that can be amortized by the execution times of the offload code.
However, in certain cases, offload code might have to be repeatedly invoked (e.g., on a series of small files/objects), and significant runtime startup latencies can be quite undesirable or even unacceptable in these situations.
Thus, we will also take a look at the startup latencies of the VMs when running our test programs.

To get the startup latencies, we would record $t_0$, the time when runtime setup starts, and $t_1$, the time when the test program can be executed, and calculate $t_1 - t_0$.
Note that this value also includes the JIT/AOT overhead.
Fortunately, this process is not that complicated, as all VMs' APIs clearly differentiate between runtime setup calls and code execution calls.

\begin{figure}[h]
  \includegraphics[width=.99\columnwidth]{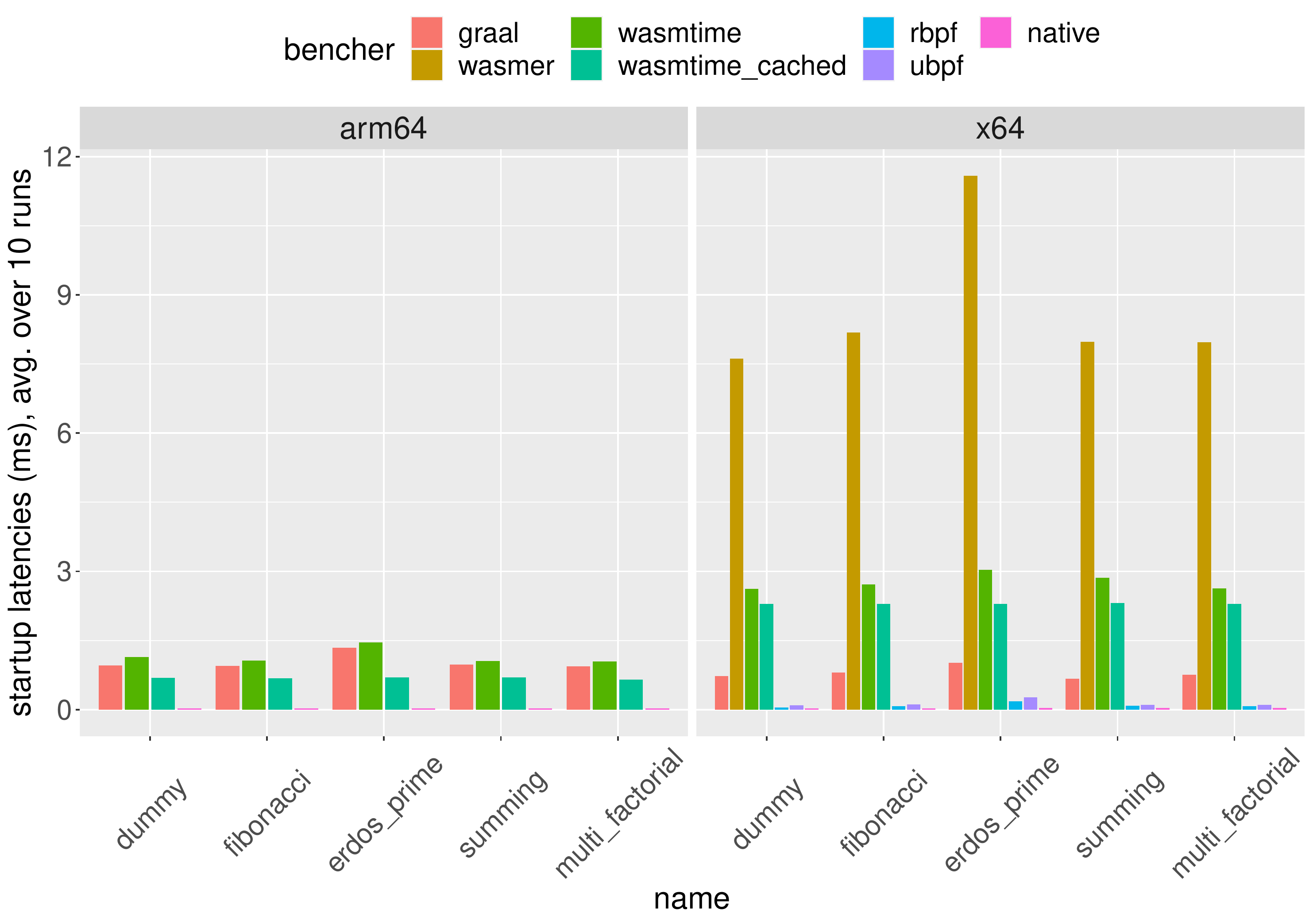}
  \vspace{-10pt}
  \caption{10-run averages of startup latencies (ms) of all \texttt{(program,VM)} combinations, absolute values}
  \label{startup_bar}
\end{figure}

\begin{figure}[h]
  \includegraphics[width=.99\columnwidth]{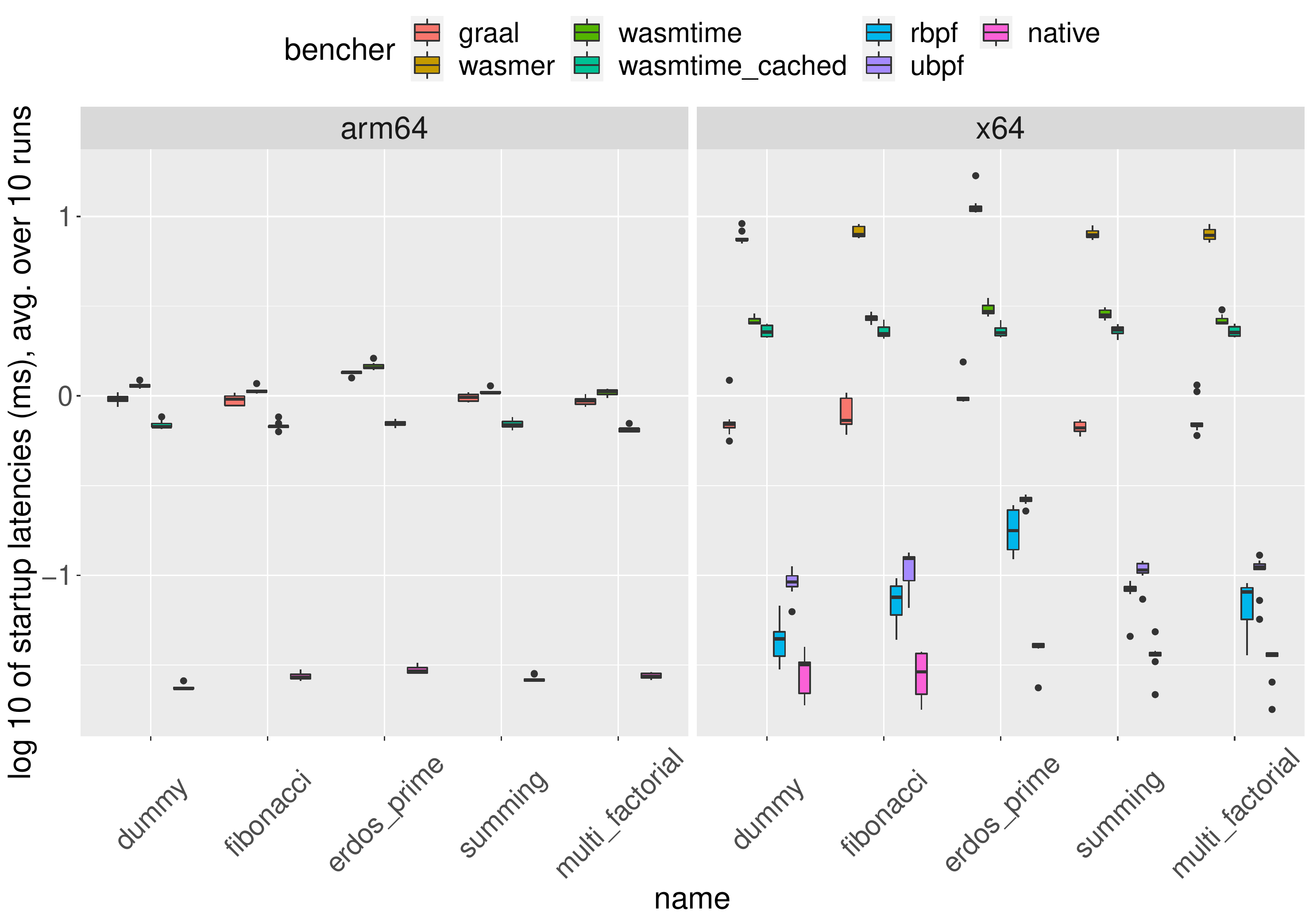}
  \vspace{-10pt}
  \caption{startup latencies (ms) of all \texttt{(program,VM)} combinations, absolute values in \texttt{log10} scale}
  \label{startup_box}
\end{figure}

From Tables \ref{x64_table} and \ref{arm64_table} (column 4) and Figures \ref{startup_bar} and \ref{startup_box}, we can see some quite dramatic variations in the startup latencies of different VMs.
As discussed before, \textit{rBPF} and \textit{uBPF} are a bit lacking in terms of their features, which means they are
likely more lightweight than the competition, so it is not surprising that they were able to perform well in this category.
On the other hand, \texttt{wasmer} was once again bogged down by the expensive AOT compilation step, making it the worst performer in this category.

That said, if one pays attention to the scale of the y-axis in Figure~\ref{startup_bar}, which shows absolute values, then one would notice that the startup latencies are all quite small -- on both architectures, they are mostly under 3 ms.
As such, they should not become a performance concern in general.

\subsection{Execution Speed}

In this section, we present the execution times of the aforementioned programs on different VMs. Here, execution time refers to the amount of time spent on actual computation (i.e., from calling the test program's entry point function to receiving its return value).

\begin{figure}[h]
  \includegraphics[width=.99\columnwidth]{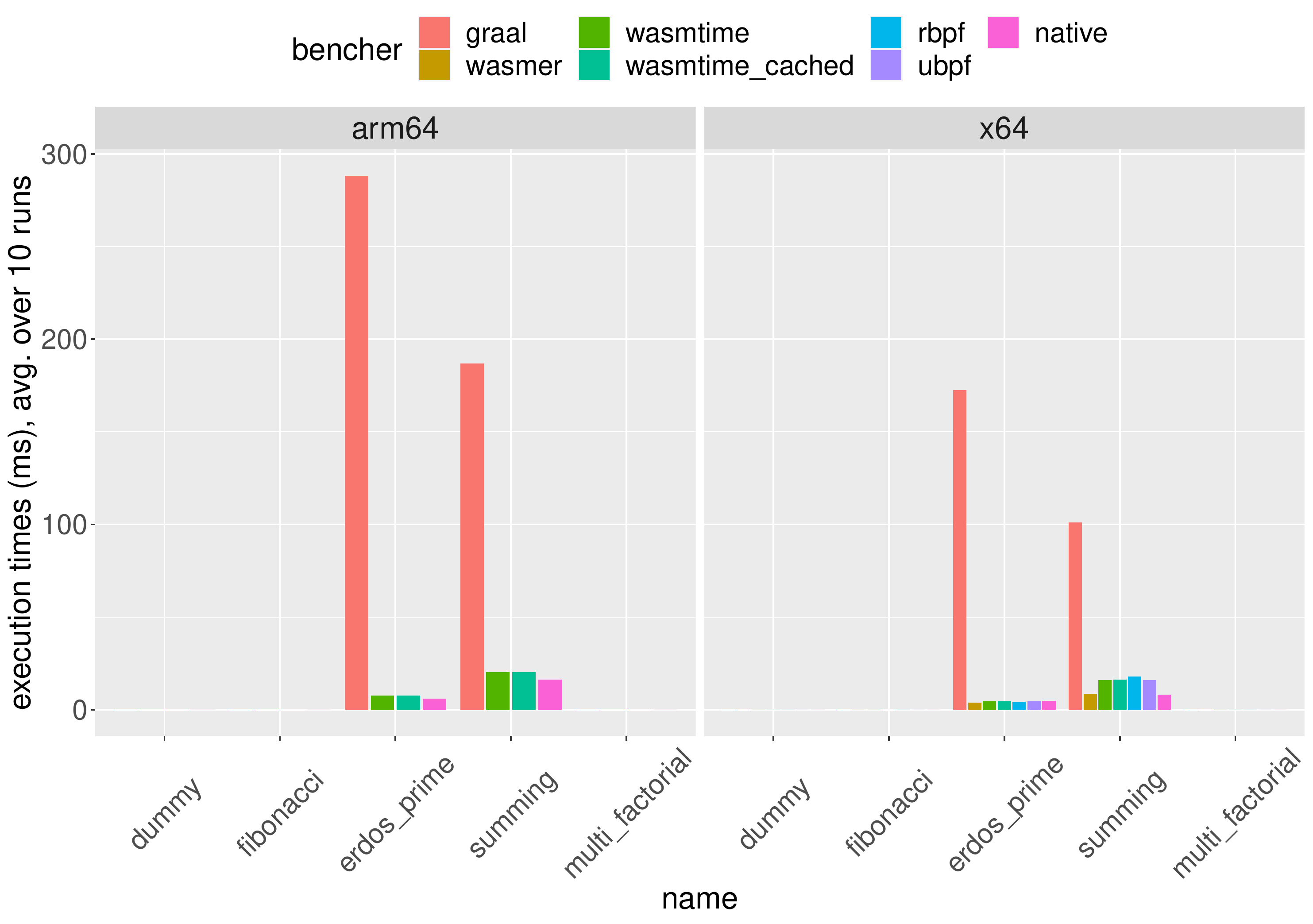}
  \vspace{-10pt}
  \caption{10-run averages of execution times (ms) of all \texttt{(program,VM)} combinations, absolute values}
  \label{execution_t_bar}
\end{figure}

\begin{figure}[h]
  \includegraphics[width=.99\columnwidth]{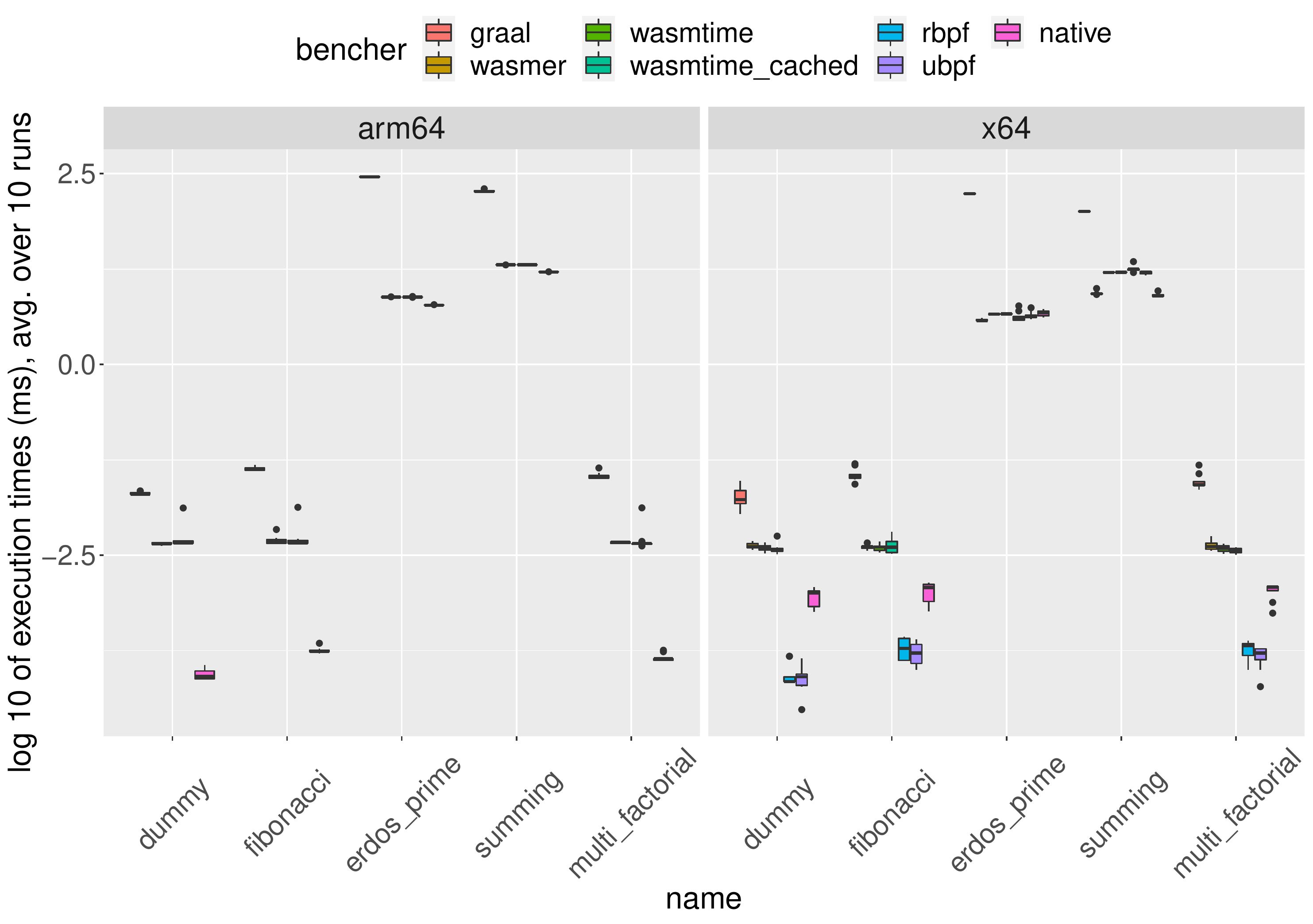}
  \vspace{-10pt}
  \caption{Execution times (ms) of all \texttt{(program,VM)} combinations, absolute values in \texttt{log10} scale}
  \label{execution_t_box}
\end{figure}

The results can be found in column 5 of Tables \ref{x64_table} and \ref{arm64_table} and in Figures \ref{execution_t_bar} and \ref{execution_t_box}.
For the execution times of the actual computation, the programs can be divided into 2 groups for discussion: (\textit{dummy}, \textit{fibonacci}, \textit{multi\_factorial}), (\textit{erdos\_prime}, and \textit{summing}).

The first group is relatively light on computation -- their bars are not even visible in Figure~\ref{execution_t_bar}.
For these programs, \textit{rBPF} and \textit{uBPF} performed so well that, amusingly, they were able to beat the native binaries by some margin.
But since these programs finished so quickly, it is not clear if we can draw any meaningful conclusion here.

The second group of programs, due to their significant usage of loops and/or memory accesses, are more intensive and can arguably put the VMs to test a little bit.
For these programs, \textit{wasmer}'s high peak memory usage and startup latencies finally paid off, as it ended up getting the best performance for \textit{erdos\_prime} and \textit{summing} on \texttt{x64}, excluding the native binaries in the latter case.
Otherwise, \textit{wasmtime}'s performance was similar to that of \textit{rBPF} and \textit{uBPF}.

There are also two more points worth mentioning.
First, on \texttt{x64}, the native binary for \textit{erdos\_prime} performed rather poorly, as it got outperformed by all VMs except \textit{GraalVM}.
Since \textit{erdos\_prime} is heavy on memory accesses, one might speculate that the VMs somehow emitted more efficient machine code in this regard.
Second, \textit{GraalVM}'s performance was pretty poor across the board, which was, again, likely due to the weight of its language interop infrastructure.

\subsection{Total Time}

In this section, we present the total running times of the aforementioned programs on different VMs.
The total running time of a test program goes from when the VM process is spawned to when the VM process exits.
Thus, it can include the time taken by the OS to perform symbol resolution and relocation, spawn the process, etc., thereby also reflecting the weightiness of the VMs a bit.

\begin{figure}[t!]
  \includegraphics[width=.99\columnwidth]{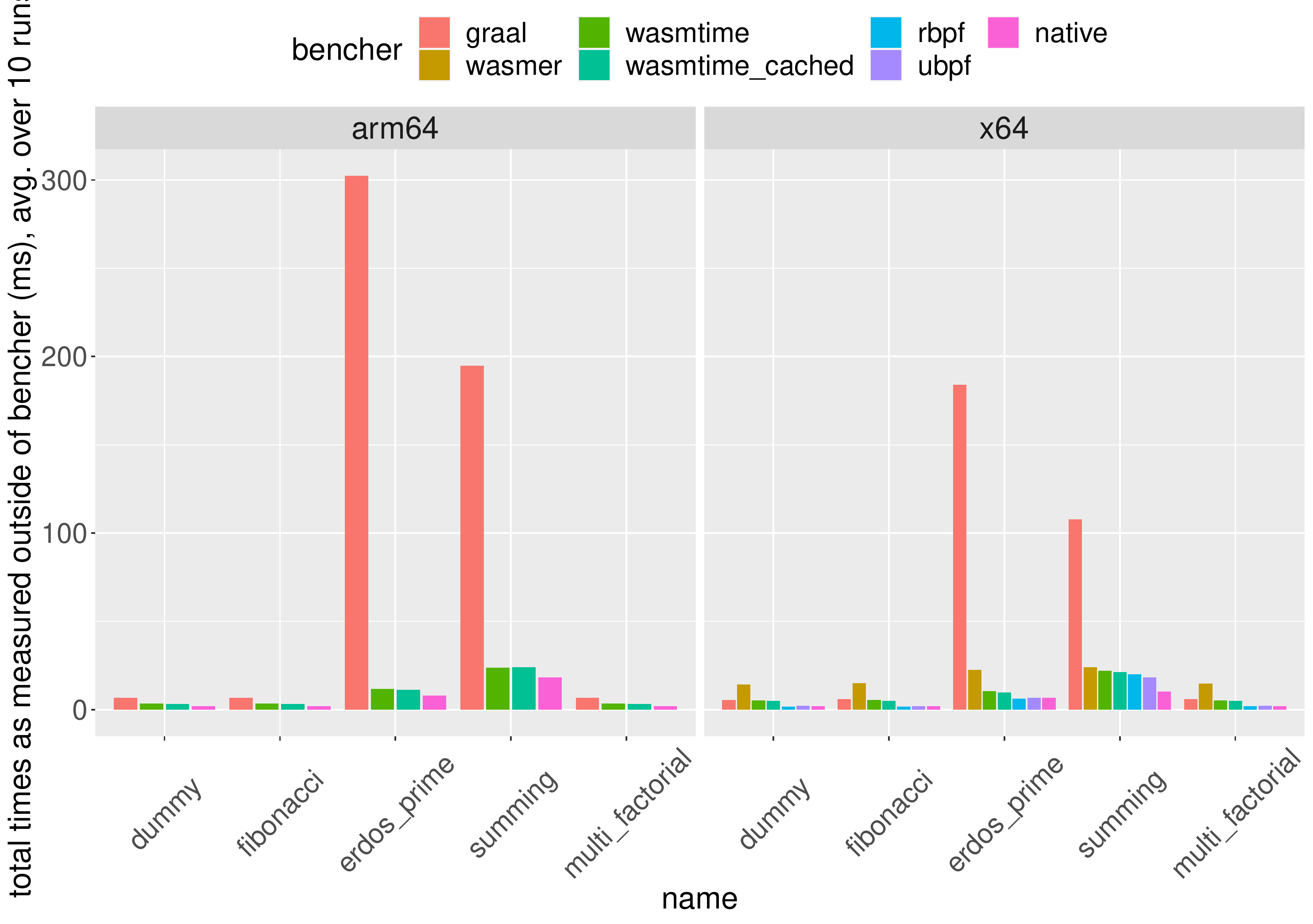}
  \vspace{-10pt}
  \caption{10-run averages of total running times (ms) of all \texttt{(program,VM)} combinations, absolute values}
  \label{total_t_bar}
\end{figure}

\begin{figure}[t!]
  \includegraphics[width=.99\columnwidth]{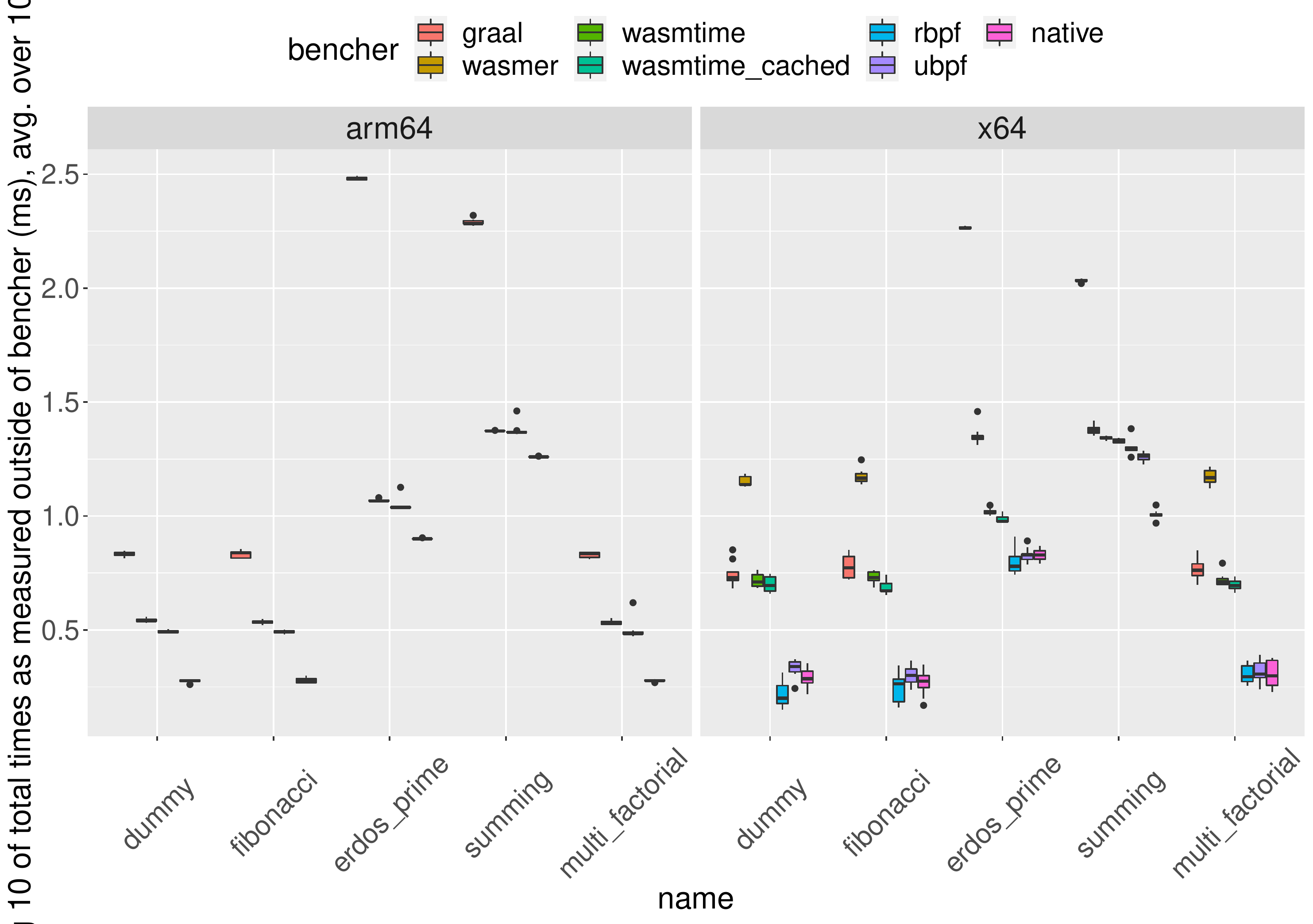}
  \vspace{-10pt}
  \caption{Total running times (ms) of all \texttt{(program,VM)} combinations, absolute values in \texttt{log10} scale}
  \label{total_t_box}
\end{figure}

Given that we collected the data in a Python script, on top of GNU \texttt{time}, the exact values might be slightly off, but since we only report the relative values in the tables above, the skew should be consistent and thus irrelevant.
The results can be found in column 6 of Tables~\ref{x64_table} and \ref{arm64_table} and in Figures \ref{total_t_bar} and \ref{total_t_box}.

\noindent
Taking everything into account, \textit{rBPF} and \textit{uBPF} come on top once again on \texttt{x64}, often beating the native binaries.
After them comes \textit{wasmtime}, whose total times are still close to those of the native binaries, especially on the \texttt{arm64} AWS instance.
Enabling caching for \textit{wasmtime} certainly helps with the total times, but the margin is not dramatic by any means.\footnote{This will of course change as the code
complexity increases.}
The AOT cost of \textit{wasmer} cannot yet be fully amortized on small programs like ours, but its numbers for \textit{erdos\_prime} and \textit{summing} are not that far behind \textit{wasmtime}'s, so we believe it has great potentials.
As for \textit{GraalVM}, once there is substantial operation involved, it is a bit too slow to compete with the other options, so unless language interop or the Java ecosystem is indispensable, we do not see why it should be used for running WebAssembly programs.\footnote{But to be fair, WebAssembly is marked as an experimental feature of \textit{GraalVM}.}

\subsection{Binary Sizes}

\begin{figure}[t!]
  \includegraphics[width=.99\columnwidth]{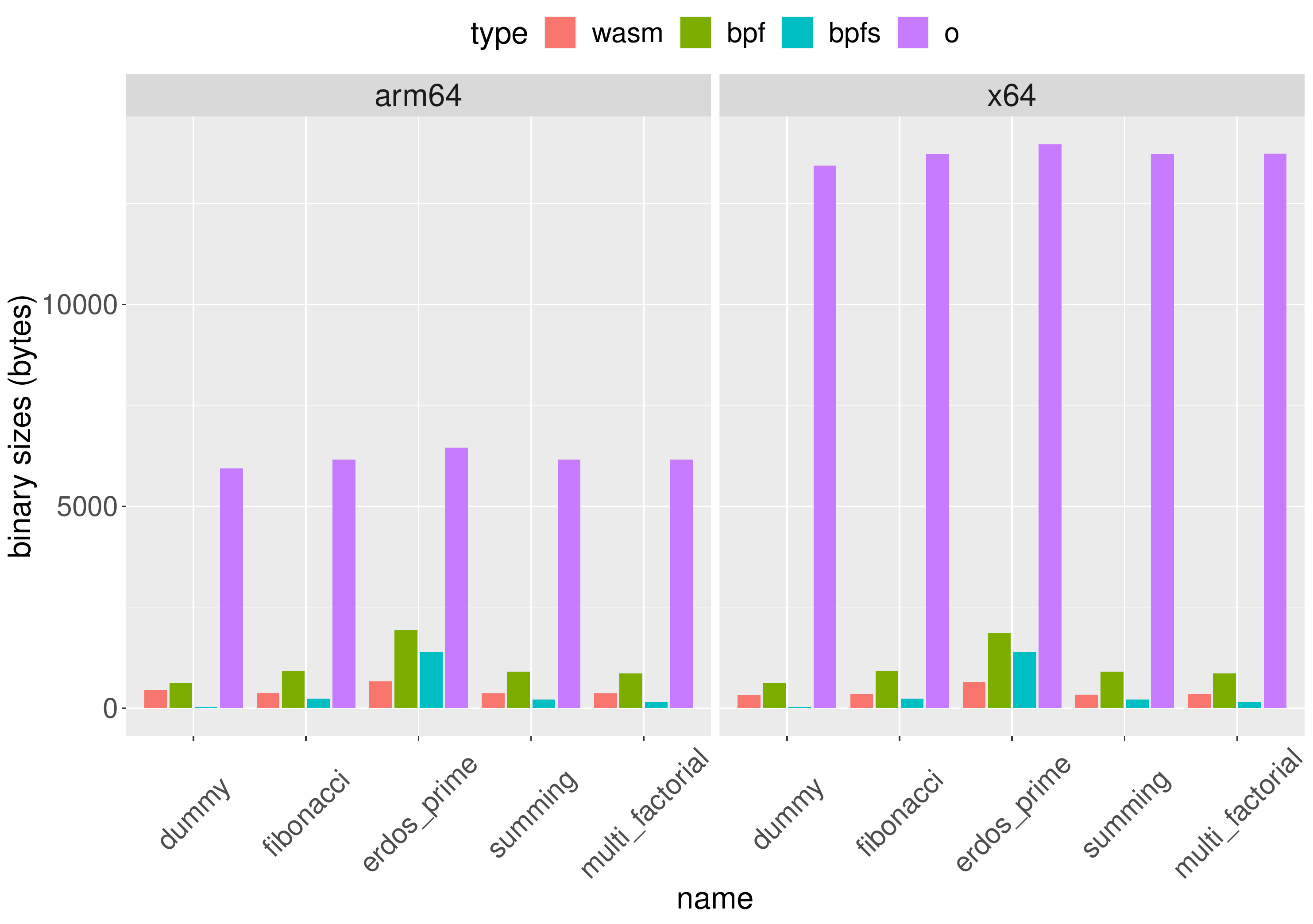}
  \vspace{-10pt}
  \caption{Binary sizes (bytes)}
  \label{binary_size_bar}
  \vspace{-15pt}
\end{figure}

The sizes of the compiled binaries can be of interest too, as over time, they can have quite some impact on the efficiency of code transmission.
The binary sizes of our test programs can be found in Table~\ref{binary_size_table} and Figure~\ref{binary_size_bar}.

\begin{table}[h]
  \begin{tabularx}{0.48\textwidth}{|*{6}{>{\centering\arraybackslash}X|}}
    \hline
    & \textbf{x64} & \textbf{arm64} & \textbf{eBPF} & \textbf{patched eBPF} & \textbf{wasm}\\
    \hline
    \textit{dummy} & \num{13440} & 5936 & 624 & 23 & 318\\
    \hline
    \textit{fibo-nacci} & 13728 & 6152 & 920 & 238 & 354\\
    \hline
    \textit{sum-ming} & 13728 & 6160 & 904 & 212 & 339\\
    \hline
    \textit{prime} & 13960 & 6456 & 1856 & 1398 & 639\\
    \hline
    \textit{multi-fact} & 13736 & 6160 & 856 & 148 & 344\\
    \hline
  \end{tabularx}
  \vspace{-10pt}
  \caption{Binary sizes (bytes)}
  \vspace{-10pt}
  \label{binary_size_table}
\end{table}

\noindent
The native binaries consistently have the largest sizes due to their inclusion of much metadata information (e.g., the \texttt{.symtab}, or symbol table, section), whereas the WebAssembly binaries consistently have the smallest sizes, if we ignore ``patched eBPF'' for a second.

\noindent
Patched eBPF ``binaries'' are a bit special.
As discussed before, these have to be manually created by a human at the moment, and they are stored as plaintext, so they are not really compiled binaries in some sense.
Nevertheless, if we take them into consideration as well, then they are indeed the smallest except for the \textit{erdos\_prime} program.

However, since \textit{erdos\_prime} also happens to be the most complicated among the programs, we would argue that, as program complexity increases, WebAssembly will likely be the winner in terms of binary sizes.

\section{Discussion}

While the eBPF VMs had a slight edge in our quantitative evaluation, the amount of restrictions and extra efforts that eBPF currently imposes on its users suggests that it is not yet ready to support data processing and general-purpose computation.
For that, WebAssembly seems to be a more sensible choice, and it is ready for deployment.
Below, we will propose some potential improvements for both technologies.

\smallskip

\noindent\textbf{WebAssembly~~}
Being a fairly mature technology, WebAssembly and its implementations can
still use some improvements in these directions:

\smallskip

\begin{itemize}[itemsep=-3pt,topsep=0pt,partopsep=0pt]
  \item stabilize and finish implementing the SIMD and threading proposals
  \item provide an interface for (preferably zero-copy) sharing of host data
  \item address the security concerns pointed out by researchers (e.g.
 ~\cite{lehmann2020everything})
\end{itemize}

\smallskip

\noindent\textbf{eBPF~~}
Following the evaluation in the previous sections, it is our opinion that eBPF still needs a lot of work before it can become a suitable vehicle for data processing:

\smallskip

\begin{itemize}[itemsep=-3pt,topsep=0pt,partopsep=0pt]
  \item a mature VM, with support for things like debugging and profiling
  \item an extended instruction set to support all common numeric operations (e.g., signed divisions)
  \item a specification that clearly defines permitted operations
  \item a reliable verifier developed in tandem with the specification
  \item implementation of \textit{libc} and other system services
  \item undergo thorough security evaluations in the context of data processing
\end{itemize}

\section{Conclusion}

In this work, we have evaluated two mechanisms for the purpose of enabling general computation offloading in the context of near-data processing -- WebAssembly and eBPF.
From our point of view, we believe that WebAssembly is a better choice for the job, but we welcome differing opinions, and we hope more discussions in these directions will ensue.

\section*{Acknowledgments}

This project has received funding from the European Union's Horizon 2020 research and innovation program under grant agreement No 957407.

\bibliographystyle{plain}
\bibliography{main.bib}

\end{document}